\documentclass[%
 reprint,
 amsmath,amssymb,
 aps,nofootinbib,   
]{revtex4-1}
\usepackage{bm}
\PassOptionsToPackage{linktocpage}{hyperref}
\usepackage[hyperindex,breaklinks]{hyperref}
\usepackage{enumitem}
\usepackage{slashed}

\renewcommand{\theta}{\vartheta}

\usepackage{array}
\usepackage{mathtools}

\usepackage{etoolbox}

\begin{document} 

\title{Absence of $\mu$-Problem in Grand Unification}

\author{Gia Dvali$^{1,2}$ and Anna Jankowsky$^{1}$}
\affiliation{%
$^1$ 
Arnold Sommerfeld Center, Ludwig-Maximilians University,  Munich, Germany, 
}%
 \affiliation{%
$^2$ 
Max-Planck-Institut for Physics, Munich, Germany
}%

\date{\today}

\begin{abstract} 
Using properties of Goldstino, we show that in generic  grand unified theories with gravity-mediated supersymmetry breaking  
the $\mu$-problem is non-existent.  What happens is that supersymmetry breaking universally induces the shifts of 
 the heavy fields that generate $\mu$ and $B_{\mu}$ terms. In the leading order, these are given by the mass of gravitino and are insensitive to 
 the scale of grand unification.  The mechanism works regardless 
 whether doublet-triplet splitting is achieved via fine-tuning or  not.  
  Moreover, we illustrate this general phenomenon 
on explicit examples of theories that achieve doublet-triplet splitting 
dynamically. These include the theories with Higgs doublet as a pseudo-Goldstone boson, as well as, the approach based on spontaneous decoupling
of the light color-triplet from quarks and leptons.  

\end{abstract}

\maketitle

\section{Introuction} 

The Hierarchy Problem in supersymmetry (SUSY) comes in form of  
two puzzles. The first one is the origin of supersymmetry breaking. The second one goes under the name of the {\it $\mu$-problem}.  
 The question is what sets - at the same scale - the following three 
unrelated mass parameters: The supersymmetric mass
of the Higgs doublet superfields in the superpotential 
\begin{equation} \label{WHH}
W = \mu H\bar{H}\,,
\end{equation}
 and the two types of the soft SUSY-breaking mass terms in the scalar potential 
 \begin{equation}\label{msoft}
 m_{soft}^2 (|H|^2 + |\bar{H}|^2) + B_{\mu}H\bar{H} + c.c.\,.  
\end{equation}
Here and thereafter, we shall denote the superfields and their 
 scalar components by the same symbols. At each occasion,  
 the meaning of the notation will be clear from the context. \\

In the present paper we shall assume that supersymmetry breaking
is due to standard gravity-mediated scenario \cite{Barbieri:1982eh}.  In this framework, 
SUSY is spontaneously broken by some 
hidden sector superfield(s) $X$ with no couplings to the 
Standard Model superfields in the superpotential.  The 
supersymmetry-breaking is communicated to the Standard Model 
fields  via the supergravity couplings. These are suppressed by the powers of the Planck mass, $M_P$.  In the observable sector, 
this mediation results in 
effective soft SUSY-breaking terms set by the scale of the gravitino mass 
$m_{3/2}$. \\ 

It has been pointed out by Giudice and Masiero \cite{Giudice:1988yz} that 
this framework offers a natural 
solution to the $\mu$-problem. 
 Namely, both $\mu$ and 
$B_{\mu}$ terms can be generated provided one postulates 
certain non-minimal 
couplings between the Higgs doublet 
and the hidden sector superfields in the K\"ahler function. \\

The goal of the present paper is to point out a distinct generic
reason for the absence of the $\mu$-problem which  
does not require an assumption of a non-minimal K\"ahler function.
  Namely, we wish to show that the $\mu$-problem 
{\it generically}  gets nullified once the Standard Model becomes 
embedded in a grand unified theory (GUT)  
with a scale $M$ much higher than $m_{3/2}$.  
  With no further efforts, such theories {\it generically} deliver:
\begin{equation} \label{main} 
\mu \sim m_{3/2} ~~ {\rm and} ~~ B_{\mu} \sim m_{3/2}^2 \,. 
\end{equation}
The dynamical mechanism behind this effect is the shift of the heavy fields 
- with masses and vacuum expectation values (VEVs) given by the scale $M$ - induced by SUSY-breaking.  This shift then delivers   
the $\mu$-term (\ref{main}), which in grand unification is protected solely by SUSY.
   \\
 
  The underlying reason for the scaling (\ref{main}) is that 
 in the limit $M_P \rightarrow \infty$ the Goldstino must reside 
 entirely within the hidden sector superfield $X$. 
  This is true,  even if $m_{3/2}$ is kept finite,  so that
 the fermion and boson masses in the observable sector stay split. 
The bottom line is that the grand unified Higgs sector 
generates (\ref{main}) irrespectively how large is 
the scale $M$. In particular, (\ref{main}) remains valid in the limit 
$M \rightarrow \infty$. \\

Note, our results are generically applicable to any extension of the 
supersymmetric Standard Model with the scale $M \gg m_{3/2}$, 
provided below $M$ the 
$\mu$-term is not forbidden by any conserved quantum number.  
An interesting thing about grand unification is that 
 this condition is universally enforced by the phenomenon of 
the {\it doublet-triplet splitting}.  \\
 
The essence of the problem is that in any GUT the Standard Model Higgs doublets 
$H, \bar{H}$ acquire the color-triplet partners 
$T,\bar{T}$ that reside within the same irreducible representation
of the grand unified group.  Because of this, the color-triplets 
can mediate an unacceptably-fast proton decay, unless some measures are taken. 
This difficulty goes under the name of the {\it doublet-triplet splitting problem}. Now the point is that, usually, the same mechanism that renders the color-triplet harmless, below the GUT scale leaves  
no protective symmetries  - other than supersymmetry (and possibly $R$-symmetry)  - for the $\mu$-term.  As a result, 
the $\mu$-term of the form (\ref{main}) is generated after SUSY-breaking due to the shifts of the VEVs of the heavy fields.  
 \\
 
 For the above reason, we shall mostly be motivated  
by grand unification but the readers can apply the present mechanism  to 
their favored high scale extensions of the SUSY Standard Model. \\

The traditional approach to doublet-triplet splitting problem is to split masses of doublets and triplets at the GUT scale. That is, upon Higgsing the grand unified group, 
the color-triplets gain the masses of order the GUT scale 
$M$, whereas the doublets remain massless.
The underlying reason for such mass splitting is model dependent.
Some examples shall be discussed below. 
 \\

In an alternative, less-traditional, approach \cite{Dvali:1992hc}, no mass splitting takes place. That is, the entire multiplet  that houses doublets and triplets remains massless at the supersymmetric level and gains a small 
mass after. So, up to higher order corrections, 
the color-triplet partner remains as light as the 
Higgs doublet. However, their couplings with quarks and leptons become 
so strongly split that the color-triplet is rendered effectively-decoupled.  
Therefore, in this scenario, the proton decay is suppressed because the 
spontaneous breaking of GUT symmetry dynamically {\it uncouples} 
the Higgs doublet's color-triplet partner from quarks and leptons
(see later). \\

  One way or another,  in GUTs,  the existence of 
massless  Higgs doublets in unbroken SUSY theory, becomes intertwined with the doublet-triplet  splitting problem. 
   The results of the present paper are largely insensitive 
to a concrete mechanism that solves this problem.  As long as the 
theory delivers a pair of massless Higgs doublets in supersymmetric 
limit, the generation of $\mu$ and $B_{\mu}$ of the form 
(\ref{main}) is generic. \\

 In fact, this way of generating $\mu$
 has been incorporated in the past 
within particular scenarios, most notably, within the 
pseudo-Goldstone approach to the doublet-triplet splitting  
problem \cite{Inoue:1985cw},\cite{Berezhiani:1989bd}  
(see also, \cite{Barbieri:1992yy, Barbieri:1993wz, Barbieri:1994kw, Berezhiani:1995sb, Csaki:1995cf, Ananthanarayan:1995tz, Dvali:1996sr, Cheng:1999fw}).  
In these scenarios, in SUSY-limit,  the Higgs doublets are
massless by Goldstone theorem.  They acquire the desired $\mu$ and 
$B_{\mu}$ terms after SUSY-breaking. 
 We shall  explain that this case represents a particular manifestation 
 of a generic shift scenario.  For this, we shall reduce the pseudo-Goldstone Higgs idea to its bare essentials and interpret the generation of the $\mu$-term in (\ref{main}) as a consequence of the Goldstino argument. 
  In fact, this case is predictive due to the interplay of Goldstino 
  and Goldstone theorems.  While the former gives 
  (\ref{main}), 
  the latter makes the relation more precise, demanding 
  \begin{equation} \label{Bmumu}
  B_{\mu} = 
  2\mu^2= 2m_{3/2}^2\,.
  \end{equation}
 This is because, by Goldstone theorem, at the tree-level,  the mass matrix of the Higgs doublets must have one zero eigenstate.
  \\

As a second example, we apply the shift mechanism to the 
 model of \cite{Dvali:1992hc} in which the light color-triplet
 is uncoupled from quark and lepton superfields. 
As said, here in  SUSY-limit the doublets and triplets are both massless. 
 We shall show that a generation of universal $\mu, B_{\mu}$ terms
 of order (\ref{main}) for both 
components takes place after SUSY-breaking.   \\

Before we move on, some comments are in order. 
First, the gravity-mediation of SUSY-breaking is important for our arguments. 
The implementation of an analogous shift mechanism in gauge mediation
requires a specific construction that has been done in \cite{Dvali:1996cu}. \\
 
 Secondly, it is important that all the singlet superfields,  
 coupled to the Higgs doublets in the superpotential, have
 large masses and that there are no {\it sliding singlets} 
 \cite{Witten:1981kv} among them.  
As it is well-known  \cite{Nilles:1982mp}, in gravity-mediation, such singlets  destabilize the weak scale.\footnote{This situation can change in 
gauge-mediated scenario with low scale of SUSY-breaking, see, 
\cite{Ciafaloni:1997gy}.}
 \\

\section{Goldstino Argument}

 We shall now present a general argument.  An impatient reader may find 
 it more useful to first go over a simple explicit example discussed in the 
 next chapter and then come back to a general proof. \\

In order to set the stage, consider a prototype supersymmetric grand unified theory which at low energies 
delivers a pair of massless Higgs doublet superfields, $H$, $\bar{H}$. 
 These  Higgs doublets are coupled in a superpotential to a singlet superfield
 $S$ with large VEV/mass.  This superfield impersonates the 
 component(s) of the heavy Higgs superfields that break the grand unified symmetry group down to  the Standard Model,  $SU(3)\times SU(2)\times U(1)$.   
 \\

 As it is customary, we assume that the 
 primary source of spontaneous breaking of supersymmetry 
 is an $F_X$-term of a (canonically normalized)  hidden sector superfield $X$. 
 This sets the absolute scale of supersymmetry breaking as 
 $M_{SUSY}^2 = \langle F_X \rangle $. In the absence of other fields, 
  the fermionic component $\Psi_X$ of the superfield $X$ is a Goldstino, 
   which becomes eaten up 
 by gravitino.  The resulting mass of gravitino is,
\begin{equation} \label{gravmass}
m_{3/2}^2 = \frac{|\langle W\rangle|^2}{M_P^4} = 
\frac{|\langle F_X \rangle|^2}{3M_P^2}\,, 
\end{equation}
where $\langle W\rangle$ is the expectation value of the superpotential.
This order parameter breaks $R$-symmetry.  The second equality in (\ref{gravmass}) is the condition for zero vacuum energy. 
 \\

Now, it is assumed that the two sectors do not talk to each other in the superpotential. 
The superpotential therefore has the following generic form: \begin{equation} \label{WHX}
W=W(S)+(\tilde{M} +gS)H\bar{H}+W(X),
\end{equation}
where $g$ is a coupling constant and $\tilde{M}$ is a mass parameter 
of order $M$ that also sets the scale in $W(S)$.    
The standard couplings of the Higgs doublets to quark and lepton 
superfields are not shown explicitly. \\

For the moment, we shall not specify the 
form of the 
superpotential $W(S)$ of the heavy 
superfield $S$.  The only assumption we make is that 
in globally supersymmetric limit it has a SUSY-preserving vacuum in which the field $S$ receives a supersymmetric mass from the VEV(s)
of the heavy field(s) given by some high scale $M$, for example, 
a grand unification scale, and that below this scale $S$ carries no conserved 
quantum number(s), with the possible exception of $R$-charge. 
However, the $R$-invariance cannot serve as an exact protective symmetry as 
it is broken together with SUSY by the non-zero gravitino mass.  
At the same time, throughout the discussion, we shall keep the VEVs of the Higgs doublets $H,\bar{H}$ zero. 
Obviously, in case of a single field $S$, with the above assumption, its mass must come from the 
self-coupling(s) in $W(S)$. We begin with this case.  
\\

Thus, in the limit of global supersymmetry 
($M_P \rightarrow \infty$, all other scales finite), we would have: 
\begin{equation} \label{I1}
\begin{split}
& \langle \frac{\partial W}{\partial S}\rangle =0 \,, ~{\rm for}  ~\,  \langle S \rangle  \equiv S_0 \sim M\,, \\
& {\rm such~that :}\,~
\langle  \frac{\partial^2 W}{\partial S^2} \rangle \sim M \,,  
\end{split} 
\end{equation} 
and 
\begin{equation} \label{I2}
0\neq\langle\frac{\partial W}{\partial X}\rangle= M^2_{SUSY}\, .
\end{equation} 

The resulting $\mu$-term is represented by the 
$S_0$-dependent supersymmetric mass of the Higgs doublets, 
\begin{equation} \label{muterm}
\mu = \tilde{M}+g S_0 \,. 
\end{equation}
We shall assume that
this $\mu$-term is zero. The underlying reason for this cancellation
is unimportant. For example, it may take place due to a dynamical reason, a group theoretic structure, or (in  a least attractive case) by fine tuning.  
Some explicit examples of cancellation mechanisms shall be discussed below. \\

 We wish to show that, regardless the nature of the cancellation mechanism, 
for finite $m_{3/2}$, the VEV of 
 the heavy singlet $S$ is shifted in such a way that the 
 $\mu$-term of order $m_{3/2}$ is generated.  We shall prove this 
 using the following Goldstino argument. \\
 
 Consider the supergravity potential for $S$ and $X$ scalar fields
 (for simplicity, we assume the minimal K\"ahler): 
 \begin{equation} \label{pot} 
 V = {\rm} e^{(|S|^2 + |X|^2)/M_P^2}\left ( |F_S|^2 + |F_X|^2 - 3m_{3/2}^2M_P^2 \right ) \,, 
  \end{equation} 
  where, 
\begin{equation} \label{FS}
F_S\equiv \frac{\partial W}{\partial S} + m_{3/2}S^* \,,
\end{equation}
\begin{equation}
F_X\equiv \frac{\partial W}{\partial X} +m_{3/2}X^*\,,
\end{equation}
and $m_{3/2} \equiv  W/M_P^2$ should be understood as
a function of the scalar components. \\
  
  Let us now consider various scaling regimes.
 First, we take $M_P\rightarrow \infty$ while keeping the scale $F_X$ finite. 
 From (\ref{gravmass}), this means that $m_{3/2} \rightarrow  0$, 
 while the product $m_{3/2}M_P$ is kept finite. We wish  to find out the scaling of $F_S$ in this regime. At first glance, this depends on the behaviour of the scale $M$.  
 For example, if we take $M/M_P \rightarrow 0$ (equivalently $m_{3/2}M \rightarrow 0$), 
 the vacuum of the $S$ superfield reduces to a globally 
 supersymmetric case (\ref{I1}), with $\langle S \rangle = S_0$ and 
$F_S = 0$. \\

 What happens if we keep $M/M_P$ finite (but, of course, small)? 
 Despite the fact that this implies $m_{3/2}M={\rm finite}$,  
 the supergravity $F_S$-term must vanish. This can be understood 
 from the following argument.   Since $m_{3/2} =0$ 
 and $M_P=\infty$, gravitino is both massless and decoupled.
  Therefore, the Goldstone fermion of spontaneously broken 
 supersymmetry must remain as a physical massless fermion. 
 By the super-Goldstone theorem, Goldstino is given by the 
following combination of $\Psi_S$ and $\Psi_X$:
\begin{equation} \label{Goldstino}
\Psi_{Gold} = \frac{\langle F_S \rangle \Psi_S+\langle  F_X \rangle \Psi_X}{\sqrt{\langle  F_S \rangle^2+\langle F_X \rangle^2}}\,.
\end{equation}
 This fermion must form a zero eigenstate of $2\times 2$ fermion mass matrix.  The second eigenvalue is of order $M$ and, therefore, is infinite
(recall that $M/M_P$  was kept finite).  
  Since the two sectors talk via supergravity, the mixing angle that diagonalizes the mass matrix, must vanish for $m_{3/2} =0$. 
  From (\ref{Goldstino}) it is clear that this mixing angle is $\sim F_S/F_X$.  We therefore conclude that  
  for $m_{3/2} = 0$, we have $F_S =0$,
 even if $M$ is taken to be infinite. 
 This indicates that for the finite values of these parameters, $F_S$ is controlled by the gravitino mass and not by the GUT scale $M$.  In particular, for $M\gg m_{3/2}\neq 0$, the $F_S$-term 
must behave as $F_S \sim m_{3/2}^2$. \\

 Let us obtain the same result more explicitly. 
 Since $F_S$ depends only on the scales $M$ and $m_{3/2}$, we can 
 compute it in power series of an infinitesimal parameter 
 $\frac{m_{3/2}}{M}$: 
 \begin{equation}\label{expansion}
  F_S = M^2 \left (c_0 + c_1\frac{m_{3/2}}{M}  
  + c_2\frac{m_{3/2}^2}{M^2} + ... \right ) \, .
  \end{equation} 
  Obviously, $c_0=0$, since, by assumption, $W(S)$ does not break 
  SUSY in the global limit.  
  The above Goldstino argument suggests 
  that $c_1=0$ as well.  Let us check this explicitly.\\ 
    
  From (\ref{pot}) it is clear that for  
 $m_{3/2} \rightarrow 0$ and $m_{3/2}M={\rm finite}$, the vacuum 
  of the $S$-superfield is determined by the condition   
  \begin{equation} \label{FS0}
F_S=\frac{\partial W}{\partial S} + m_{3/2}S^* = 0 \,.
\end{equation}
  Notice that this condition does {\it not} reduce to a globally supersymmetric condition  (\ref{I1}). This is because $S_0 \sim M$
  and, therefore, the term $m_{3/2}S^*$ cannot be ignored. 
  Let us find the solution to the condition (\ref{FS0}) in form of a  
small shift around the globally supersymmetric value: 
$S= S_0 + \delta S$.  Plugging this in (\ref{FS0}), we obtain that the  
cancellation of the leading terms demands, 
   \begin{equation} \label{FSD}
\left (\frac{\partial^2W}{\partial S^2}\right )_{S=S_0} \delta S + m_{3/2}S_0^* = 0 \,.
\end{equation}
Since  $\left (\frac{\partial^2W}{\partial S^2}\right )_{S=S_0} \sim 
S_0 \sim M$, we have 
\begin{equation} \label{ShiftS}
\delta S \sim m_{3/2} \,.
 \end{equation} 
 
 Thus, we have shown that the shift of a field that gets the supersymmetric mass 
 from its large VEV, is of order $m_{3/2}$.  We observe that  
 this shift, in the leading order, is independent of the scales $M$ and $M_P$. \\

  This result implies that, if the $\mu$-term (\ref{muterm}) is zero in 
  supersymmetric limit, after SUSY-breaking,  
   $\mu \sim m_{3/2}$ is generated. 
  \\
  
  The generalization of the above reasoning for arbitrary number of 
  heavy superfields  $S_j,~j=1,2,...$, is straightforward. It is convenient 
  to work in the eigenstate basis of the global-SUSY mass matrix 
  $M^{ij} \equiv \frac{\partial^2W}{\partial S_i\partial S_j}|_{S_j = S_{0j}}$, 
  where, as previously, $S_{0j}$ denote the globally supersymmetric 
  VEVs of the superfields that satisfy  
  $\frac{\partial W}{\partial S_j} =0$.
  Such a basis always exists since the matrix $M^{ij}$
  is holomorphic and 
  symmetric. The equation (\ref{FSD}) then becomes, 
     \begin{equation} \label{FSD}
M_{jj}  \delta S_j + m_{3/2}S_{0j}^* \, = \, 0 \,.
\end{equation}
Notice, this is a leading order relation that ignores unimportant 
contributions of order $m_{3/2}^2$.  
Now, since by our starting assumption, $M_{jj} \sim M$ for all $j$-s and 
$S_{0j} \sim M$ at least for some $j$-s, the corresponding 
shifts are $\delta S_j \sim m_{3/2}$.

\section{Simple Example}   
  
We can illustrate the action of the above general 
mechanism for an explicit form of the 
superpotential: 
\begin{equation} \label{WS}
W(S)=\frac{M}{2}S^2+\frac{\lambda}{3}S^3,
\end{equation}
where $M$ is a high scale and $\lambda$ is a coupling constant 
of order one. 
In the global SUSY limit, the VEV of $S$ is given  by 
\begin{equation}\label{Szero}  
\frac{\partial W(S)}{\partial S}=MS+\lambda S^2=0
 \rightarrow S_0= -\frac{M}{\lambda}.
\end{equation}  
We wish to determine the shift $\delta S$ triggered by 
the gravity-mediated supersymmetry breaking. 
From the Goldstino argument presented above, this shift can be found from the condition (\ref{FS0}) which in the present case translates as 
\begin{equation}
 MS + \lambda S^2  \simeq  - m_{3/2} S^* \, .
\end{equation}
This gives the shift
$\delta S = - \frac{m_{3/2}}{\lambda}$. 
The explicit minimization of the entire potential including the standard 
gravity-mediated soft terms,
\begin{equation}
\begin{split}
V = &|\lambda S + MS|^2 + m_{3/2}^2|S|^2 +\\
+& \left (m_{3/2}A\frac{\lambda}{3} S^3 + m_{3/2}(A-1)\frac{M}{2}S^2 + c.c. \right ) \,,
\end{split}
\end{equation} 
fully confirms this result. \\

 Now, we recall that in global supersymmetry limit 
 the $\mu$-term was assumed to be zero. That is, 
 we have, 
 \begin{equation} \label{effmu}  
 \mu =\tilde{M} + g S_0 =  \tilde{M} - \frac{g}{\lambda} M =0 \,. 
 \end{equation}
  Then, 
 the shift of $S$, induced by the soft terms, 
 generates a non-zero $\mu$-term given by  
\begin{equation} \label{Smu} 
\mu = 
g\delta S =  -\frac{g}{\lambda}m_{3/2} \,.
\end{equation}

 To conclude this section, we have argued that the generation of the 
 $\mu$-term of order $m_{3/2}$, due to the shift of the VEV(s) of  heavy field(s), is  rather generic.  This phenomenon is independent of a precise 
 mechanism that sets $\mu=0$ in the supersymmetric limit.  
  
 \section{Decoupling} 
 
  The Goldstino argument indicates that it is in general wrong 
  to work with the low energy superpotentials obtained by 
  substitution of the supersymmetric VEVs of the heavy fields.
   Indeed, in the above example,  the  
   observable sector described by the superpotential, 
   \begin{equation} \label{DEC} 
   W = \frac{M}{2}S^2+\frac{\lambda}{3}S^3 + 
   (\tilde{M} +gS)H\bar{H} \,,
   \end{equation}
   had a supersymmetric vacuum
   \begin{equation}
   S = S_0= -\frac{M}{\lambda}\,,~~H=\bar{H} = 0\,.   
   \end{equation}
   In this vacuum, the superfield $S$ had a mass $M$, whereas 
   the mass of the doublets (\ref{effmu}) was fine-tuned to zero.\\
   
    Now, since the mass of the superfield $S$ is much higher than the 
    supersymmetry breaking scale, it may be tempting to integrate 
    $S$ out while ignoring supersymmetry breaking. This naive 
  approach  would 
    give an effective low energy superpotential for the 
 doublets $H,\bar{H}$ with zero $\mu$-term plus 
 high dimensional operators suppressed by the scale $M$.  
 Then, neglecting the high dimensional operators, one would arrive 
 to a low energy theory with $\mu=0$. 
   Such a description would 
 completely miss the generation of the $\mu$-term due to the 
 shift of the heavy field $S$.  \\
 
This may seem a bit confusing, since we expect that 
the effects of the heavy fields must be suppressed by powers of 
their mass $M$.  However, obviously, 
there is no conflict with 
the principle of decoupling. 
 What is happening in reality is that, although 
 the mass term of the heavy scalar $S$ scales as 
 $\sim M^2$, so does the tadpole generated by supersymmetry breaking, 
 which goes as $\sim m_{3/2}M^2 S$. As a result, 
 the shift  $\delta S \sim m_{3/2}$ is finite 
 even in the limit $M\rightarrow \infty$.  This effect must be taken into account when integrating out the heavy fields properly. \\
 
 \section{GUTs}
 
 \subsection{Example with fine tuning: $SU(5)$}   
 
 The essence of how the dynamical generation of the $\mu$-term  
 is intertwined with doublet-triplet splitting, 
 can be illustrated on a prototype GUT example of minimal $SU(5)$. 
 As it is well known, in this theory the Higgs doublets 
 $H, \bar{H}$ and their triplet partners $T,\bar{T}$ are embedded 
 in $5_H, \bar{5}_H$ representations respectively. The 
 breaking of $SU(5)$ symmetry down to the Standard Model 
 group is achieved by the $24_H$ Higgs representation. 
 The superpotential of the Higgs superfields is:  
 \begin{equation} \label{SU5}
W= \frac{M}{60} tr24_H^2 -  \frac{\lambda}{90} tr24_H^3
+ \tilde{M} \bar{5}_H5_H - \frac{g}{3} \bar{5}_H24_H5_H \,.
\end{equation}
Note, it is the necessity of the doublet-triplet splitting that excludes the possibility of setting $\tilde{M}$ and $g$ small. This eliminates  
any symmetry protection for the resulting $\mu$-term which - after 
GUT symmetry breaking -
is left solely at the mercy of SUSY.  
 \\  

 Substituting into (\ref{SU5}) the only non-zero component $24_H=S{\rm diag}(2,2,2,-3,-3)$, the system effectively reduces 
 to the example (\ref{WS}) (and (\ref{DEC})) with an extra pair of $T,\bar{T}$ superfields,   
 \begin{equation} \label{5EFF}
W(S)=\frac{M}{2}S^2+\frac{\lambda}{3}S^3 
+ (\tilde{M}  + g S) \bar{H}H + (\tilde{M} -\frac{2}{3} g S) \bar{T}T \,.
\end{equation}
 The supersymmetric VEV of $S$ is given by (\ref{Szero})
and the doublet-triplet splitting is achieved by fine-tuning 
(\ref{effmu}). This fine-tuning gives the zero $\mu$-term for the doublets
and simultaneously generates a large supersymmetric mass term for their color-triplet partners, 
$\mu_T = \frac{5g}{3\lambda} M$. This takes care of the suppression
of proton decay. \\

Now, as already explained, after SUSY-breaking the $\mu$-term 
of order $m_{3/2}$ is generated due to 
the shift of the VEV of the $S$-superfield and is given by 
(\ref{Smu}).  \\

One may argue that, in theory with fine tuning, we are not gaining 
much by inducing the required $\mu$-term after SUSY-breaking.  
After all, it is not clear why the fine tuning $\mu=0$ is any more natural  
than a fine tuning to order TeV.   
One could try to dispute this by saying that
for the superpotential (\ref{SU5})  - which knows nothing about the weak scale - the scales $M$ and zero are the two natural points. \\

However, we shall not do this. By default, such disputes usually take one 
to nowhere due to the lack of the guiding principle in theories 
with fine tuning.  This is why we are more attracted to scenarios 
in which $\mu=0$ in SUSY limit is justified by the underlying 
structure of the GUT theory.  \\

However, there is an important point that works regardless of 
fine tuning:  The value of $\mu$  in the low energy theory is shifted 
by 
\begin{equation} \label{mushift} 
\delta \mu \sim m_{3/2}\,,
\end{equation}
 with respect to its SUSY value.  This exposes an intrinsic 
sensitivity of the supersymmetric Standard Model towards 
the GUT-completion. \\

   This concludes the example with fine tuning. In what follows, we shall illustrate the same effect 
 on examples of theories which, in unbroken supersymmetry, 
 achieve the vanishing $\mu$ dynamically.

 \subsection{Example: Higgs as a Pseudo-Goldstone} 
  
  As the first example, we consider class of theories 
  in which the Higgs doublets $H,\bar{H}$ are 
 pseudo-Goldstone bosons \cite{Inoue:1985cw},\cite{Berezhiani:1989bd}. In these models, 
 before supersymmetry breaking, the $\mu$-term is dynamically adjusted to zero by the Goldstone theorem. \\
  
 The idea is that 
 the {\it Higgs part} of the GUT superpotential has large accidental global symmetry.  This global symmetry is spontaneously broken at  the 
 GUT scale along with the local one.  
 This breaking results into a pair of 
 (pseudo)Goldstone superfields with the quantum numbers of $H,\bar{H}$.  
  Due to the global symmetry of the Higgs part of the superpotential,  
  in supersymmetric limit, these superfields are exactly massless. 
 Correspondingly, before SUSY-breaking, $\mu=0$.
  After the soft SUSY-breaking terms are included, 
 the $\mu \sim m_{3/2}$ is generated. In the minimal case
 (with canonical K\"ahler metric), due to 
 Goldstone theorem, at the tree level, one combination of 
 doublets $H,\bar{H}$ remains massless even after supersymmetry breaking. 
 This degree of freedom acquires a non-zero mass and a VEV via  radiative corrections.  \\
  
  The above idea was realized in two main directions, \cite{Inoue:1985cw} and \cite{Berezhiani:1989bd}, 
  where \cite{Berezhiani:1989bd} represents a justification of \cite{Inoue:1985cw} from a more fundamental theory. We shall briefly discuss the key aspects of the two approaches. 
    \\  
  
  The first proposal was a model  by Inoue, Kakuto and Takano  and by Anselm and Johansen \cite{Inoue:1985cw}. Both examples were based on a minimal  $SU(5)$ GUT.  As already discussed, the Higgs sector of this theory consists 
 of an adjoint $24_H$-plet and a pair of $5_H,\bar{5}_H$-plet chiral superfields. This theory exhibits a textbook example of the doublet-triplet 
 splitting problem; The required mass-splitting between the
 color-triplet and the weak doublet components of $5_H,\bar{5}_H$ is achieved 
 at the expense of a severe fine tuning discussed in the previous chapter. 
 \\ 
 
 The idea by the authors of \cite{Inoue:1985cw} was that this fine-tuning admits 
 an interpretation in terms of the Goldstone theorem, provided 
 an additional gauge-singlet chiral superfield, $1_H$,  is added  
 to the Higgs sector.  In such a case, after careful adjustment 
 of the parameters, the Higgs part of the superpotential 
 becomes invariant under a global $SU(6)$ symmetry group. 
 Under it, various $SU(5)$ Higgs multiplets 
 combine into a single adjoint representation: 
 $35_H = 24_H + 5_H + \bar{5}_H + 1_H$. In such a case, the 
 Higgsing of the gauge $SU(5)$ symmetry is accompanied 
 by a spontaneous breaking of the global $SU(6)$ symmetry.
 The latter breaking results into left-over pseudo-Goldstone multiplets 
 $H,\bar{H}$ with the quantum numbers of the electroweak Higgs 
 doublets. These are the doublet components of  $5_H,\bar{5}_H$ respectively. \\
   
  The potential criticism against this scenario is that a severe fine tuning  
among two large numbers is traded for more severe fine tunings  
among several large parameters. 
In order to dissolve this criticism, one needs to justify the demanded 
global $SU(6)$ pseudo-symmetry as an {\it accidental} symmetry emerging 
from a more fundamental theory.  This was achieved 
in \cite{Berezhiani:1989bd}
   by lifting (i.e., UV-completing) the theory into a GUT with 
 a  gauged $SU(6)$ symmetry.  The accidental global symmetry then 
 emerges as a low energy {\it remnant} of this gauge symmetry.   
 This happens in the following way. \\

The minimal set of chiral superfields necessary for Higgsing the $SU(6)$ gauge symmetry down to the Standard Model group, $SU(3)_c \times SU(2)_L \times U(1)_Y$, consists of an adjoint $35_H$-plet 
and a 
pair of $\bar{6}_H, 6_H$-plets. 
If the cross coupling $\bar{6}_H 35_H 6_H$ is absent,  the 
renormalizable superpotential of the Higgs fields splits 
into two non-interacting parts, 
\begin{equation}\label{WU6}  
W_H = W(35) + W(6_H,\bar{6}_H)\,.  
\end{equation}
Obviously, this superpotential 
has a global symmetry $SU(6)_{35_H} \times SU(6)_{6_H}$
under independent $SU(6)$-transformations of the two sectors.   
The subscripts indicate the superfields on which the corresponding symmetries act.  \\

Without entering into a discussion about naturalness, we 
note that the global symmetry $SU(6)_{35_H} \times SU(6)_{6_H}$ can be viewed as 
accidental. This is because it results from the absence of a single cross-coupling, 
as opposed to fine tuned cancellations among several  
big numbers. In addition, there have been proposals 
of justifying the absence of this cross coupling from more fundamental 
theory, such as, for example, a stringy anomalous $U(1)$ symmetry \cite{Dvali:1996sr} or  locality in the extra space \cite{Cheng:1999fw}. \\

The Higgsing of the gauge symmetry down to Standard Model group 
is triggered by the following vacuum expectation values (VEVs):
\begin{equation} \label{VEV}
\langle 35_H \rangle = \textrm{diag}(1,1,1,1,-2,-2)v_{35} , ~~
 \langle  6_H \rangle = \langle  \bar{6}_H \rangle = \begin{pmatrix}
v_6\\
0\\
0\\
0\\
0\\
0
\end{pmatrix},   
\end{equation}
where the parameters $v_{35}$ and $v_6$ are of order GUT scale. \\

Now, simultaneously with the Higgsing 
of the gauge group, the global symmetries are spontaneously broken 
in the following way, 
\begin{subequations} \label{SU(4)}
\begin{align}
SU(6)_{35_H} & \rightarrow SU(4) \times SU(2) \times U(1)\\
SU(6)_{6_H} &\rightarrow SU(5)
\end{align}
\end{subequations}
The straightforward count of the Goldstone bosons and
the diagonalization of the mass matrix 
 shows that one pair 
of chiral superfields, with the quantum numbers of electroweak 
doublets, is left-out ``uneaten" by the gauge fields and remains exactly massless. 
These superfields are the two linear combinations of the doublets 
$H_{35}, \bar{H}_{35}$ and $H_{6}, \bar{H}_{\bar{6}}$ from the $35_H$ and $6_H,\bar{6}_H$ fields respectively:  
\begin{equation} \label{uneaten}
H = \frac{H_{35}v_{6} - 3 H_{6}v_{35}}{\sqrt{v^2_{6} + 9v^2_{35}}}, ~~ \bar{H} = \frac{ \bar{H}_{35}v_{6} - 3 \bar{H}_{\bar{6}}v_{35}}{\sqrt{v^2_{6} + 9v^2_{35}}}
\end{equation}
At the same time, all colored components of the Higgs superfields acquire masses of order the GUT scale.
 Thus, the doublet-triplet splitting is achieved as a result of 
the Goldstone phenomenon. The two Higgs doublets (\ref{uneaten}) 
which represent the Goldstone modes of the accidental global symmetry, are strictly massless in the limit of exact SUSY.
Correspondingly, the $\mu$-term vanishes in this limit.  \\

The $\mu = m_{3/2}$ is generated after supersymmetry breaking. 
In the previous analysis this was demonstrated by an explicit minimization 
of the Higgs potential in the presence of the soft SUSY-breaking terms.
   Our goal here is to view the generation of 
   the $\mu$-term in the pseudo-Goldstone model 
 as a manifestation of the universal shift mechanism discussed in the present paper.  We shall therefore reduce the Goldstone mechanism 
 to its bare essentials. \\

The unified property of such models is that the 
Higgs doublets $H,\bar{H}$ are connected - by a continuous global (pseudo) symmetry 
- to some GUT superfield(s) ($N,\bar{N}$)
with large VEV(s). The latter fields Higgs the gauge GUT
symmetry and simultaneously break spontaneously the
continuous global symmetry that connects them with $H,\bar{H}$.     
For all practical purposes, we can effectively characterize the above  
global degeneracy as a $U(3)$ global symmetry 
in which the gauge $SU(2)_L\times U(1)_Y$ electroweak symmetry enters as $U(2)$ subgroup. We must note that this simplified formulation of the pseudo-Goldstone mechanism was previously proposed in \cite{Dvali:1996cu}. \\

Naturally, under this $U(3)$, the $U(2)$-doublets $H,\bar{H}$ and 
singlets $N, \bar{N}$ form the triplet and anti-triplet representations
respectively:
  \[3_H=\begin{pmatrix}
H \\ N
\end{pmatrix}, \bar{3}_H=\begin{pmatrix}
\bar{H} \\ \bar{N}
\end{pmatrix} \,. \] 
 The precise form of a $U(3)$-invariant superpotential that leads to the desired symmetry breaking, $U(3) \rightarrow U(2)$, is not important. For simplicity, we can choose it 
 as:
\begin{equation} \label{super}
W = \lambda S(3_H\bar{3}_H - M^2),
\end{equation}
where $S$ is a singlet superfield, $\lambda$ is coupling constant and $M$ is a scale
of symmetry breaking. For simplicity, we take all parameters real. 
\\

It is easy to check that the above superpotential gives a globally supersymmetric vacuum in which the continuous global symmetry 
is broken down to Standard Model $U(2)$ by the following VEVs 
\[ \langle N \rangle = \langle \bar{N} \rangle = M.\]  At the same time, 
the VEVs of 
all other superfields are zero, $S=H=\bar{H}=0$. 
 Obviously, the $\mu$-term is also zero, $\mu = \lambda S= 0$.  
This happens because the doublets $H, \bar{H}$ represent the Goldstone bosons 
of spontaneously broken global symmetry $U(3) \rightarrow 
U(2)$.
\\

 After the soft SUSY-breaking terms are included, 
 the VEVs get shifted. These shifts induce nontrivial mass terms in the Higgs sector and both the $\mu$- and $B_{\mu}$- terms are generated.\\

In order to find them, we minimize the potential 
with soft supersymmetry breaking terms included:
\begin{equation}
\begin{split}
V= &|\lambda (H\bar{H} + N\bar{N} - M^2)|^2 + \\
+ & |\lambda S|^2 (|H|^2 + |\bar{H}|^2 + |N|^2 + |\bar{N}|^2)+\\
+& m_{3/2}^2(|S|^2  + |H|^2 + |\bar{H}|^2 + |N|^2 + |\bar{N}|^2) + \\
+& m_{3/2}A\lambda S(H\bar{H} + N\bar{N}) + m_{3/2}(2-A)\lambda M^2S + c.c. \,,
\end{split}
\end{equation}
After a straightforward calculation, we obtain that,  in the leading order in expansion in small parameter $\frac{m_{3/2}}{M}$,  the VEVs
are given by:
\begin{equation}
\begin{split}
&\langle N \rangle = \langle \bar{N} \rangle  = M+\frac{m_{3/2}^2}{2\lambda^2 M} (A-2)\,, \\
& 
\langle S \rangle = -\frac{1}{\lambda}m_{3/2}.
\end{split}
\end{equation}

Next we insert the above VEVs  in the Lagrangian relevant for the masses of $H,\bar{H}$:
\begin{equation}
\begin{split}
\mathcal{L}\supset& (|\lambda S|^2 + m_{3/2}^2) (|H|^2 + |\bar{H}|^2) +\\
 &
+ \left ((|\lambda|^2 (N\bar{N}-M^2)^* 
+ Am_{3/2}\lambda S)H\bar{H} + c.c. \right) \\
\equiv& (\mu^2+m_{3/2}^2) (|H|^2 + |\bar{H}|^2) + ( B_{\mu}H\bar{H}
+ c.c.) \,,
\end{split}
\end{equation}
where, 
\begin{equation}\label{Bmumu1}
B_{\mu} = - 2 m_{3/2}^2 \,,~ 
 \mu = - m_{3/2}\,, 
\end{equation}
in accordance to (\ref{Bmumu}). \\

Hence, the $\mu$- and $B_{\mu}$- terms are produced at the same scale
and the resulting mass matrix has the form:
\begin{equation} \label{mass2}
\hat{M}^2_H=\bordermatrix{
	& H^*	& \bar{H}	\cr
H	& 2m_{3/2}^2	&- 2m_{3/2}^2	\cr
\bar{H}^*	& - 2m_{3/2}^2	& 2m_{3/2}^2	\cr
}
\end{equation}
Notice, due to the existence of a Goldstone mode at the tree-level, 
the mass matrix \eqref{mass2} has a zero-eigenvalue. 
The latter equality is specific to the pseudo-Goldstone approach.
However, the universal feature shared by other approaches is that   
the gravity-mediated SUSY-breaking generates 
$B_{\mu}$ and  $\mu$ at the scale set by $m_{3/2}$,
as given by (\ref{main}).

\subsection{Example: Decoupled Triplet}

  The last example in which we shall implement the generation of the $\mu$-term by the shift mechanism, is the approach to doublet-triplet splitting problem developed in \cite{Dvali:1992hc}. 
   In this picture the weak doublets ($H, \bar{H}$) as well as their 
   color-triplet partners ($T, \bar{T}$), are isolated from 
   the VEVs that break GUT symmetry down to the Standard Model. 
   As a result, all these superfields remain exactly massless in supersymmetric theory. That is, no mass splitting among the doublets and triplets takes place. Instead,  the entire GUT multiplet remains massless. \\
   
 This may come as a surprise, since it is expected that 
 light color-triplets $T,\bar{T}$ mediate proton decay at unacceptable 
 rate. 
 However, in the scenario of \cite{Dvali:1992hc}
   this potential problem is avoided
 by decoupling the color-triplets $T,\bar{T}$ 
 from the quark and lepton superfields. 
Only the doublets $H,\bar{H}$ maintain the usual coupling to quarks and leptons. 
To put it shortly, in this scenario the doublet-triplet splitting 
gets transported from the mass terms into the 
Yukawa couplings.
 As a result, the proton decay is equally strongly suppressed both at 
 $d=6$ and $d=5$ operator levels. 
 \\

  Let us, following \cite{Dvali:1992hc}, consider a realization of this idea in 
  a supersymmetric  
  $SO(10)$ theory in which the Higgs doublet resides in $10_H$ representation.  
The quarks and leptons are placed in $16_F$ spinor representation. 
 The idea is that $10_H$ couples to matter fermions 
 via an intermediate heavy $45_H$ Higgs that has a VEV on 
 $SO(6)\times SU(2)\times U(1)$-invariant component: 
 \begin{equation} \label{45VEV}
 \langle 45_H\rangle = M_{45} \, {\rm diag}(0,0,0,\epsilon,\epsilon) \,,~~
 \epsilon \equiv \begin{pmatrix}
    0  & 1   \\
    -1  &  0
\end{pmatrix}\,,
 \end{equation}
 where $M_{45}$ is of order GUT scale.  
 The coupling with matter fermions is generated by the exchange of a pair
of  heavy $144, \overline{144}$-dimensional multiplets with the following couplings 
 in the superpotential:
 \begin{equation}
 W_{\rm F} = g 16\gamma_i\overline{144}_j 45_{Hij} + M_{144} 
 144_j\overline{144}_j + g'10_{Hj} 144_j16 \,,
 \end{equation}
 where $j = 1,2, ..., 10$ is $SO(10)$ tensor index, 
 $\gamma_j$ are $SO(10)$ gamma matrixes and spinor indexes are not shown explicitly. $M_{144}$ is a mass term of order the grand unification scale and $g,g'$ are dimensionless coupling constants.
  The integration-out of  the $144$-dimensional multiplets results in the following 
 effective coupling in the superpotential\footnote{For illustrative 
 purposes, here we are only concerned with the 
 minimal structure of the theory. The generation of realistic fermion masses, as 
 usual, requires the enrichment of the horizontal structure which can be incorporated without changing any of our conclusions 
 and shall not be attempted here.}, 
 \begin{equation} \label{Y10}
 W_{\rm F}\,  \rightarrow  \, \frac{gg'}{M_{144}}10_{Hi} 45_{Hij} 16_F \gamma_j 16_F \,.   
 \end{equation} 
  Taking into account the form (\ref{45VEV}) of the VEV of $45_H$, it is clear that 
 the electroweak doublet components $H,\bar{H}$
 of $10_H$ acquire the usual 
 Yukawa couplings with the Standard Model fermions, 
 given by $\frac{gg' M_{45}}{M_{144}}$. 
  At the same time, 
 their color-triplet partners, $T,\bar{T}$, decouple.
  \\
 
 In this way, the burden of generating the huge masses for 
 the color-triplets, while keeping their weak-doublet partners massless, 
 is avoided.  The entire $10_H$ multiplet can be kept massless
 in supersymmetric limit.   
 All one needs for achieving this, is to assume that solely the heavy fields 
 with zero VEVs interact directly with $10_H$-plet in the superpotential. 
 \\
 
 Let such a superfield be $S$. At the same time, $S$ is free to
 (and in general will) interact with the Higgs multiplets that participate in 
 the breaking of $SO(10)$-symmetry.  Then, according to our arguments, 
 SUSY-breaking shall result into the shift of the VEV $S \sim m_{3/2}$
 and in the subsequent generation of the $\mu$-term. \\
 
  An example of the superpotential that validates this 
 mechanism is  
\begin{equation} \label{10H}
W =  S(\lambda 10_H^2 + \lambda' N^2 - M^2),
\end{equation}
where $N$ impersonates the heavy superfield(s)
that Higgs the $SO(10)$-symmetry. $M$ is a mass scale 
and $\lambda$ and $\lambda'$ are coupling constants. For 
definiteness, we take all parameters to be real and positive. \\

Of course, for achieving the right symmetry breaking 
pattern a lot more terms and a garden variety of representations 
are required. This is the standard  ``engineering" problem 
in $SO(10)$ GUT and is not specific to the present discussion. 
Our goal here is not in a construction of a fully functioning $SO(10)$ theory but rather in 
pointing out an universal shift mechanism for the 
$\mu$-term. We shall therefore focus on (\ref{10H}). 
For a detailed analysis of the Higgs sector leading to a desired symmetry breaking patterns, the reader is referred to \cite{Dvali:1995hp}. 
\\  
 
  Now, in supersymmetric limit we have $N = M$
 and $S=10_H =0$.  
 Therefore, the $\mu$-term that is set by the VEV of $S$ is zero
 and the entire $10_H$-plet is massless. 
 It is straightforward to check that the soft supersymmetry breaking generates the shift 
 $S= - \frac{m_{3/2}}{2\lambda'}$.  Correspondingly, the $\mu$-term  generated as a result of this shift 
 is $\mu=  - \frac{\lambda}{2\lambda'}m_{3/2}$. \\
   
Notice, the same $\mu$-term is generated for the color-triplet partners
$T, \bar{T}$ since they share the $10_H$ multiplet with the Higgs doublets.
Of course, the exact doublet-triplet mass degeneracy shall be lifted by radiative and other higher order corrections, but the color-triplets
shall remain light. Because these  particles are essentially decoupled from the light fermions, they are extremely long-lived. The existence of the long-lived colored 
multiplets, with their masses correlated with the Higgs doublets, is  
a prediction of the decoupled triplet scenario \cite{Dvali:1992hc}. \\

  This latest topic gives us an opportunity to observe another 
crucial impact on the low energy physics from the shift of the heavy VEVs. 
 Namely, without such a shift, the colored triplets $T, \bar{T}$ would remain 
 decoupled from the Standard Model fermions. As a result, they would remain stable.  However, the SUSY-breaking shifts the VEV of 
 the heavy $45_H$-plet and generates the entries $\sim m_{3/2}$ 
 in three empty $2\times 2$ diagonal blocks of (\ref{45VEV})\footnote{This can be easily checked by a straightforward explicit computation for simple superpotentials for various choices of $SO(10)$ Higgs multiplets that in global SUSY limit deliver the VEV (\ref{45VEV}).}.  
  This component of the $45_H$ VEV breaks 
 the $SO(6)$-symmetry down to 
 $SU(3)\times U(1)$. Simultaneously, via (\ref{Y10}),  
 this generates the effective couplings of the color-triplets 
 $T,\bar{T}$ to the quark and lepton superfields. 
 From (\ref{Y10}) it is clear that the 
 resulting decay constant 
of each color-triplet is by a factor $\sim \frac{m_{3/2}}{M_{45}}$ smaller as compared to its doublet partner Higgs.  
For $m_{3/2} \sim$TeV and $M_{45} \sim 10^{16}$GeV,  
the resulting decay-time of a color-triplet into Standard Model Particles  
is $\tau \sim$ sec or so. 
Such a long-lived colored state has potentially-interesting collider 
signatures \cite{Dvali:1992hc}, \cite{Barbieri:1992yy}, \cite{Dvali:1995hp},
\cite{Dvali:1996hs},  \cite{Berezhiani:2011uu}.

\section{Conclusions} 

The purpose of the present paper was to provide evidence that 
in a large class of grand unified theories the $\mu$-problem
is non-existent. What happens is the following. The doublet-triplet splitting mechanism that in supersymmetric theory delivers a pair of massless Higgs doublets,
after soft supersymmetry-breaking  generates the required $\mu$ and  $B_{\mu}$ terms.  This effect is independent of how the splitting was achieved in the first place. For example, the splitting can be  
arranged by a direct fine-tuning or via a more natural dynamical mechanism. 
We gave some general arguments, in particular based on Goldstino   
theorem, showing that the shifts in VEVs of the heavy fields 
universally result in generating the $\mu$-term of order $m_{3/2}$.  \\

After giving general arguments, we have shown how the 
mechanism works both in a fine tuned scenario as well as 
in two illustrative examples with dynamical
solutions to the doublet-triplet splitting problem.  \\

In the first class of theories \cite{Inoue:1985cw},\cite{Berezhiani:1989bd} 
 the Standard Model Higgs
doublet is a pseudo-Goldstone boson. 
 In this scenario, the generation of the required $\mu$ and $B_{\mu}$ 
 terms from the shift of the heavy VEVs,
 can be understood from the combination of Goldstone and Goldstino theorems. 
 The first theorem demands that the mass matrix of the Higgs doublets  
has one exact zero eigenstate at the tree-level.  At the same time 
the Goldstino argument presented here requires  that the entries 
are of order $m_{3/2}^2$.  An explicit computation for the minimal K\"ahler 
confirms this and gives the relation (\ref{Bmumu}). 
\\

Our second example is based on the scenario of \cite{Dvali:1992hc}
in which 
no mass splitting takes place between the Higgs doublet and its color-triplet
partner.  Instead, in the supersymmetric limit both components remain 
exactly massless. 
The proton decay is nevertheless safely suppressed because the
color-triplet is decoupled from quark and lepton superfields. In this scenario too, 
after supersymmetry breaking, the universal $\mu$-term is generated 
both for the Higgs doublets and their color-triplet partners.   The theory therefore predicts 
the existence of long lived colored states with their masses correlated 
with the masses of the Higgs doublets.   \\

 This scenario illustrates another important low energy effect 
 originating from the shift of the VEVs of the heavy Higgses. Namely, this shift is the sole source 
 for generating the non-zero decay constant of 
 the color-triplet partners of the Higgs doublets. 
 As a result, these color-triplets acquire the finite, albeit macroscopically long, life-times.  This makes them into the subjects of phenomenological 
 and cosmological interests. \\

Due to its generic nature, the presented shift mechanism is expected to generate the $\mu$-term in other group-theoretic solutions of the doublet-triplet splitting problem. In particular, we have explicitly checked this  
 by constructing a simple realization of Dimopoulos-Wilczeck mechanism
 \cite{DW}.
  The details will be given elsewhere. 
  The similar phenomenon is expected to work 
  also for the Missing Partner mechanism
  \cite{Georgi:1981vf, Masiero:1982fe, Grinstein:1982um}. \\

 Also note that, although we mostly focused here on dynamical scenarios, 
 the models that achieve doublet-triplet splitting via an explicit fine-tuning
 (such as the minimal $SU(5)$, considered above)
 should not be left out.  An interesting example is provided by a predictive $SO(10)$-theory of \cite{Aulakh:2003kg}. The universal shift mechanism for generating the $\mu$-term discussed in present 
 paper  shall be equally operative in such scenarios. \\

 The described phenomenon once again teaches us an important lesson that the heavy fields must be integrated out only after
  the effects of SUSY-breaking on their VEVs are properly taken 
  into account.  In the opposite case, an important impact on the low energy 
  physics from the high energy sector could be overlooked. 
  What we observe is that the $\mu$-term of the supersymmetric 
  standard model is directly sensitive to the 
SUSY-breaking-induced shifts of the heavy VEVs.  It is therefore  meaningless 
to talk about the $\mu$-problem, the least, without  knowing the
details of UV-theory. \\

In grand unification, the Higgs doublets are unified with color-triplet 
partners in the same multiplet and share quantum numbers with them.   
Then, the mandatory doublet-triplet splitting
usually strips the $\mu$-term of all unbroken quantum numbers that could potentially forbid 
its generation. As a result,  the SUSY-breaking  generically 
induces the $\mu$-term by the dynamical shift mechanism 
discussed in this paper. 
  \\

  \section*{Acknowledgements}

Part of the content of this paper has been presented in the lectures on supersymmetry delivered at NYU and were shared with David Pirtskhalava.  We thank him for discussions and involvement at the preliminary stage. 
We are grateful to Goran Senjanovic for valuable discussions and comments.
 
This work was supported in part by the Humboldt Foundation under Humboldt Professorship Award, by the Deutsche Forschungsgemeinschaft (DFG, German Research Foundation) under Germany's Excellence Strategy - EXC-2111 - 390814868,
and Germany's Excellence Strategy  under Excellence Cluster Origins.

\end{document}